\UseRawInputEncoding
\documentclass[aps,prl,superscriptaddress,reprint]{revtex4-1}
\usepackage{amsmath}
\usepackage{amssymb}
\usepackage{graphicx}
\usepackage[colorlinks,urlcolor=blue]{hyperref}
\usepackage{url}
\usepackage{color,xcolor}
\usepackage{ulem}
\usepackage{float}
\usepackage{epstopdf}
\begin{document}

\title{Orbital-selective Peierls phase in the metallic dimerized chain MoOCl$_2$}
\author{Yang Zhang}
\author{Ling-Fang Lin}
\affiliation{Department of Physics and Astronomy, University of Tennessee, Knoxville, TN 37996, USA}
\author{Adriana Moreo}
\author{Elbio Dagotto}
\affiliation{Department of Physics and Astronomy, University of Tennessee, Knoxville, TN 37996, USA}
\affiliation{Materials Science and Technology Division, Oak Ridge National Laboratory, Oak Ridge, TN 37831, USA}

\date{\today}

\begin{abstract}
Using {\it ab initio} density functional theory, here we systematically study the monolayer MoOCl$_2$ with a $4d^2$ electronic configuration. Our main results is that an orbital-selective Peierls phase (OSPP) develops in MoOCl$_2$, resulting in the dimerization of the Mo chain along the $b$-axis. Specifically, the Mo-$d_{xy}$ orbitals form robust molecular-orbital states inducing localized $d_{xy}$ singlet dimers, while the Mo-$d_{xz/yz}$ orbitals remain delocalized and itinerant. Our study shows that MoOCl$_2$ is globally metallic, with the Mo-$d_{xy}$ orbital bonding-antibonding splittings opening a gap and the Mo-$d_{xz/yz}$ orbitals contributing to the metallic conductivity. Overall, the results resemble the recently much discussed orbital-selective Mott phase but
with the localized band induced by a Peierls distortion instead of Hubbard interactions.
Finally, we also qualitatively discuss the possibility of OSPP in the $3d^2$ configuration, as in CrOCl$_2$.

\end{abstract}

\maketitle

\textit{Introduction.}
Transition-metal (TM) compounds continue attracting the attention of the Condensed Matter community~\cite{Dagotto:rmp94,Dagotto:rp,Kotliar:pd,Dagotto:science05,Dagotto:Rmp}. In these materials a wide variety of interesting phenomena have been found driven by the Coulomb repulsion $U$ and Hund coupling $J_H$. The list includes high-$T_c$ superconductivity~\cite{Bednorz:Cu,Dagotto:Rmp,Kamihara:Jacs,Dai:Np,Li:Nature,Zhang:prb20}, orbital ordering~\cite{Tokura:science,Pandey:prb,Lin:prm21}, multiferroicity due to charge or spin ordering~\cite{Brink:jpcm,Lin:prm17,Dong:nsr,Zhang:prb20-2}, and orbital-selective Mott phases (OSMP)~\cite{OSMP,Patel:osmp,Herbrych:osmp1,Herbrych:osmp2}. Interestingly for our purposes, molecular-orbital clusters are also known to exist in some inorganic compounds containing correlated electronic TM atoms~\cite{Streltsov:Phys.-Usp,Khomskii:cr}, such as dimers in Li$_2$RuO$_3$~\cite{Miura:JPSJ} and (TaSe$_4$)$_2$I~\cite{Zhang:prb20-1}, trimers in Ba$_3$Ru$_4$O$_{\rm 10}$~\cite{Streltsov:prb12}, heptamers in AlV$_2$O$_4$~\cite{Horibe:prl} and others~\cite{Haverkort:prl,Kim:prb14,Streltsovt:prb14,Lin:prb21,Kimber:prb14,Korotin:sr}.

\begin{figure}
\centering
\includegraphics[width=0.48\textwidth]{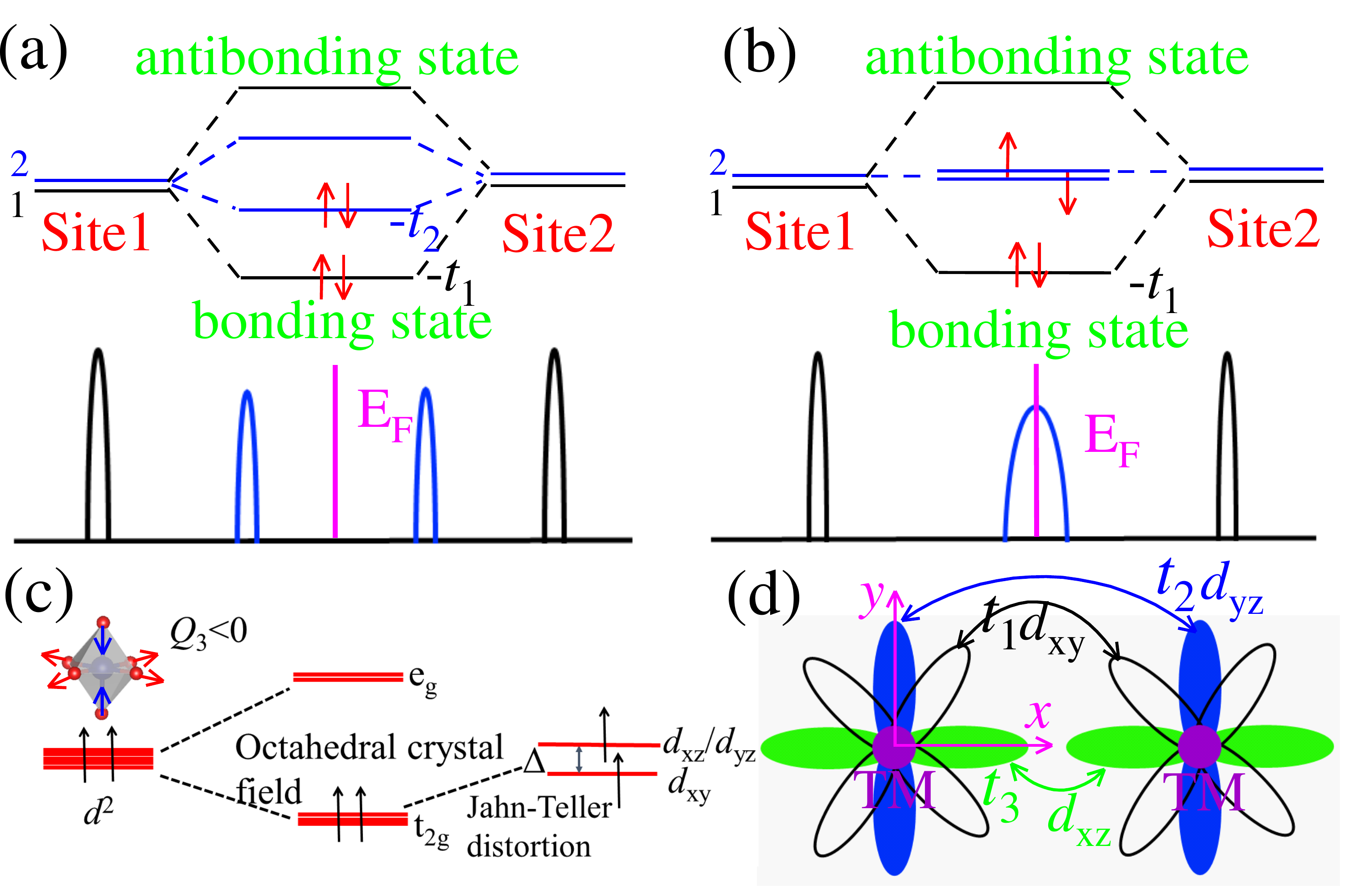}
\includegraphics[width=0.48\textwidth]{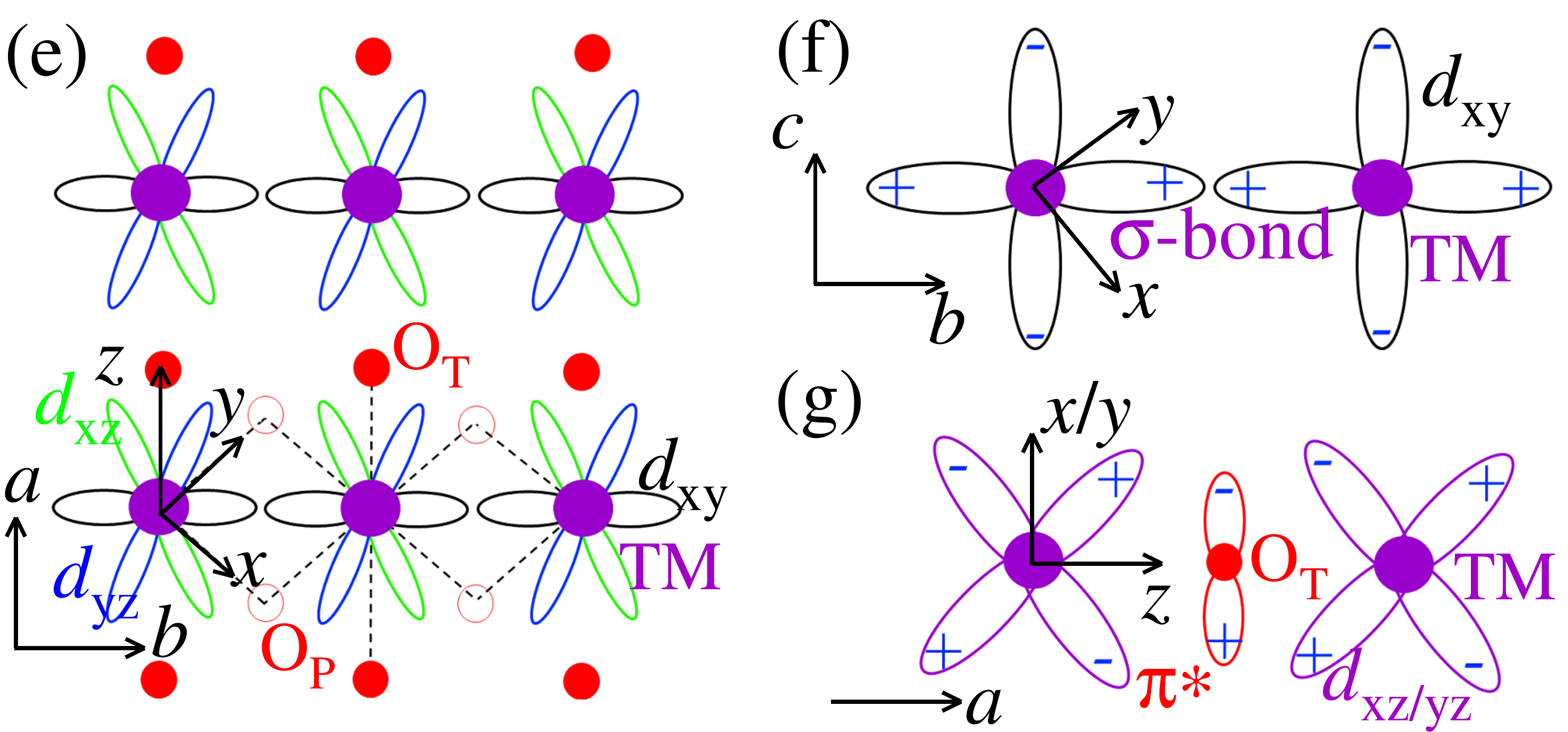}
\caption{(a-b) Two possible electronic configurations in a dimer. Here, we use two electrons in two orbitals per site, as an example. (a) Both orbitals form bonding-antibonding molecular-orbital states, resulting in an insulating phase (the sketched local density-of-states is localized). (b) The $\gamma = 1$ orbital forms strong molecular orbitals, while $\gamma = 2$ remains decoupled (the sketched local density-of-states has coexisting localized and itinerant features). (c) Schematic energy splitting of TM's $d$ orbitals with the $d^2$ configuration ($Q_3 \textless 0$) in the presence of the Hund rule. (d) Schematic three hoppings in one dimer for the $d^2$ configuration when three $t_{2g}$ orbitals are active.  (e-g) Possible lattice structures for real materials in the OSPP metallic phase. O$_T$ and O$_P$ indicate the oxygen atoms at the top or in plane of the TMO$_6$ octahedron.Using the same level of detail of panel (d) would complicate panel (e) unnecessarily, thus in (e) each orbital is only sketched in a qualitative manner. (e) Two-dimensional sketch of a possible crystal structure with the TM-O-TM chain along the $a$-axis and TMO$_2$ chain along the $b$-axis, respectively. (f) Sketch of a TMO$_2$ chain. Due to the strong overlap of $d_{xy}$ orbitals along the $b$-axis, it could form a singlet dimer with $dd$$\sigma$-bonding. (g) Sketch of a TM-O-TM chain. The $d_{xz/yz}$ and O-$2p$ orbitals form a $\pi^*$ state along the $a$-axis, inducing metallic conductivity along this axis. For better clarity, we changed the lattice vectors of the conventional TMO$_6$ octahedral structure to the lattice axes of MoOCl$_2$.}
\label{Fig1}
\end{figure}

Consider the tight-binding portion of the Hamiltonian in a multiorbital system with the atomic distances dimerized, first without the electronic correlations incorporated. If the intradimer hoppings of all orbitals are much larger than the interdimer hoppings, this system is in a global insulating phase due to the molecular orbital (MO) states formed and associated bonding-antibonding splitting of all the orbitals [see Fig.~\ref{Fig1}(a)]. However, if only one of those orbitals has a large intraorbital hopping due to strong orbital overlap, while the other orbitals' hoppings are smaller, the system could remain metallic, as shown in Fig.~\ref{Fig1}(b). Case (b) is the OSPP of our focus in this publication.

Previous work showed that some materials could be in an OSPP state, such as  Li$_2$RuO$_3$~\cite{Miura:JPSJ}, Y$_5$Mo$_2$O$_{12}$~\cite{Streltsov:jmmm}, and CuIr$_2$S$_4$~\cite{Bozin:nc}. However, by introducing electronic correlations, the molecular-orbital states could be suppressed if the Hund coupling $J_H$ is larger than the intrahopping of the MO state, or if the orbitals with much smaller hoppings $t$ would be localized due to the electronic correlation $U$ ($U \gg t$), leading to an insulating phase~\cite{Pchelkina:prb,Torardi:jssc,Bozin:nc}. Orbital-selective Peierls systems that are also metallic are rare, especially when involving molecular-like dimers. The rutile MoO$_2$, with a three-dimensional complex crystal structure, is an example of an orbital-selective Peierls metallic phase with Mo-Mo dimers, where some electrons form singlet dimers and other electrons form metallic bands~\cite{Eyert:jpcm}. Thus, the key qualitative question we will address is whether MoOCl$_2$ can be discussed
in a similar context, even after incorporating correlation effects. For this purpose, we will use density functional theory (DFT) calculations.

Experimentally, MoOCl$_2$ was found to be a strongly correlated dimerized metal based on $T$-dependent transport measurements~\cite{Wang:prm20}, and here we provide an explanation for its metallic behavior associated with the existence of an orbital-selective Peierls metallic phase. This phase exists in a regime of robust Hubbard $U$, thus it can be considered strongly correlated electrons. Our main result is that the OSPP transition indeed occurs along the $b$-axis in this system, resulting in a Mo-dimerized chain. Specifically, the Mo-$d_{xy}$ electrons form singlet dimers that are decoupled from each other along the $b$-axis, leading to a large bonding-antibonding splitting. In addition, the other occupied Mo-$d_{xz/yz}$ orbitals display strong anisotropy in the electronic structure along the $a$-axis and provide the metallic conductivity. Thus, the global metallic nature of MoOCl$_2$ is induced by the orbital-selective behavior. Furthermore, we also qualitatively discuss the possibility of orbital-selective singlet in this system but involving $3d$ atoms. We found the OSPP is suppressed by enhanced electronic correlations and reduced intradimer hopping, leading instead to the orbital-selective Mott phase for multi-orbital $3d$ TM atoms.

\textit{Model system.}
The simplest case is a $d^2$ system with a TMO$_6$ octahedral structure (Jahn Teller distortion $Q_3 \textless 0$)~\cite{Q3context}, where one electron is in the $d_{xy}$ orbital and another electron is shared between the $d_{xz}$ and $d_{yz}$ orbitals, as shown in Figs.~\ref{Fig1}(c) and (d). For possible ``real'' material OSPP realizations, the TM octahedral should be edge-sharing connected, providing the possibility of strong overlap of $d_{xy}$ orbitals along the $xy$ plane. If the edge-sharing TM octahedral could form  one-dimensional chain geometric structures along one direction in the $xy$ plane [see Fig.~\ref{Fig1}(e)], the $xy$ orbital will display strong anisotropy. Thus, the  $xy$-orbital MO singlet state will form in this direction, as shown in Fig.~\ref{Fig1}(f). Then, the other orbitals will have smaller hoppings in this direction ($t_1 \gg t_2/t_3$) if the other orbitals also display anisotropy along other directions [see Figs.~\ref{Fig1}(e) and (g)]. Hence, the Peierls transition via the $d_{xy}$ orbital, with other orbitals displaying itinerant metallic behavior, can dominante as long as the hopping strength can compete with the electronic correlation. If this is the case, then an OSPP metallic phase can be obtained in this unique system even with strong electronic correlations. Does any real material realize this physics?

\begin{figure}
\centering
\includegraphics[width=0.48\textwidth]{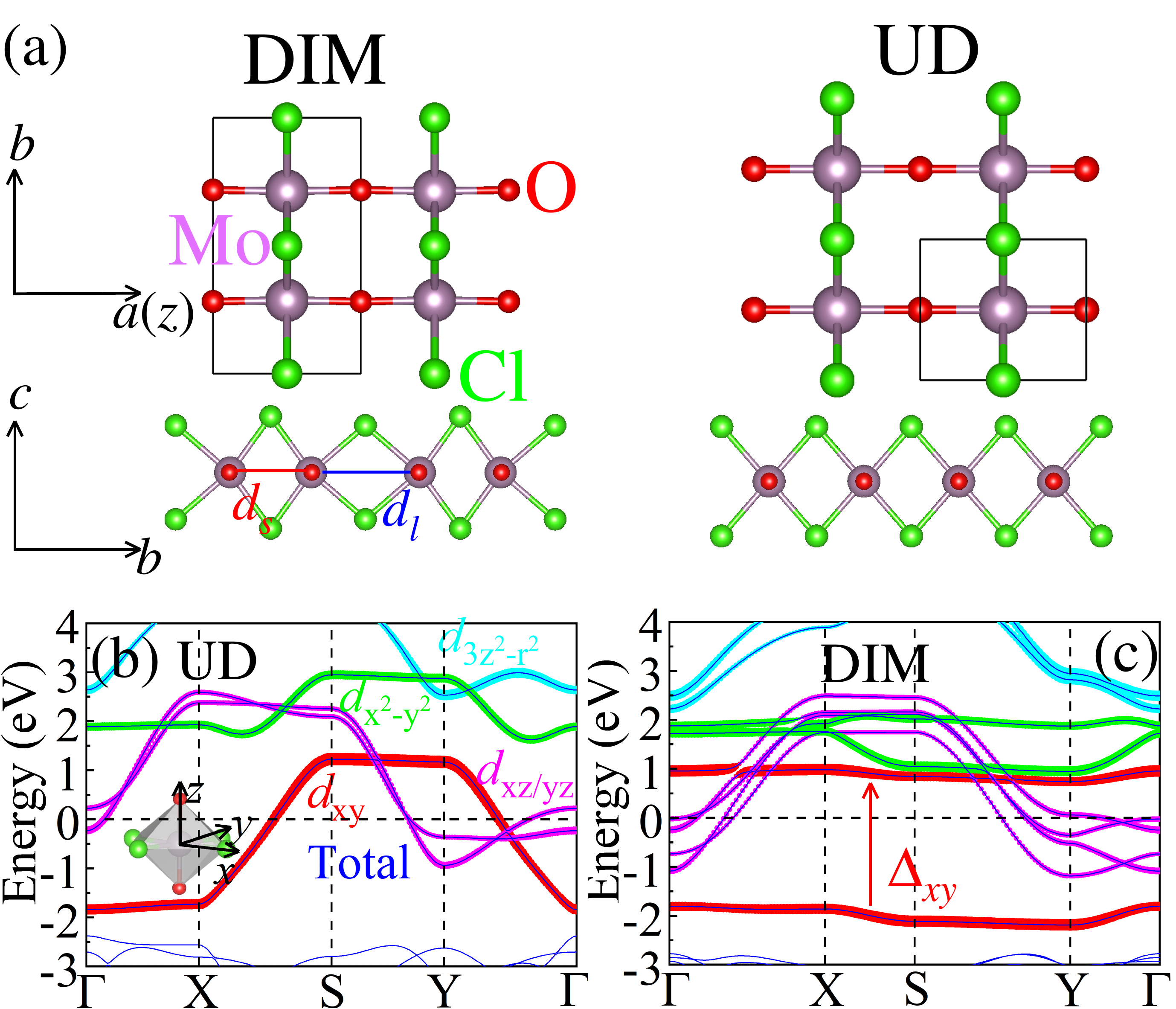}
\caption{(a) Schematic crystal structure of the MoOCl$_2$ conventional cell (magenta = Mo; green = Cl; red = O) for dimerized and undistorted phases, respectively. (b-c) Projected band structures of the monolayer MoOCl$_2$ for the non-magnetic state: in (b) we use the undistorted phase (without Mo-Mo dimer) and in (c) the dimerized phase (with Mo-Mo dimer). The Fermi level is shown with dashed lines. The weight of each molybdenum orbital is represented by the circles' sizes. The coordinates of the high-symmetry points in the plane Brillouin zone are  $\Gamma$ = (0, 0, 0), X = (0.5, 0, 0), S = (0.5, 0.5, 0), Y = (0, 0.5, 0).}
\label{Fig2}
\end{figure}

The van der Waals (vdW) family of layered oxide dichlorides $M$O$X_2$ ($M$ = V, Ta, Nb, Os; $X$ = Halogen element)~\cite{Hillebrecht:jac,Schnering:ac} is known to display the required geometric structure, where the $M$O$_2$$X_4$ octahedra are corner-sharing along the $a$-axis, while edge-sharing along the $b$-axis. Peierls distortions have been found along the $b$-axis~\cite{Ruck:acc,Jia:nh,Zhang:prb21}, resulting in $d_{xy}$ molecular orbitals, but the $d_{xz/yz}$ orbitals are unoccupied because this material is in the $d^1$ configuration. Then, it only partially fulfills our requirements for OSPP. However, recently molybdenum oxide dichloride (MoOCl$_2$) with the desired $4d^2$ configuration was reported to be a strongly correlated metal~\cite{Wang:prm20} with Mo-Mo dimers along the $b$-axis [see Fig.~\ref{Fig2}(a)]. In this compound, the replacement of Cl by O at the octahedral apex could increase the crystal-field splitting between $d_{xy}$ and $d_{xz/yz}$ orbitals, likely inducing a strong molecular-orbital $d_{xy}$ dimer. Furthermore, in the $4d$ periodic table row the reduced electronic correlation strength may allow for $d_{xz/yz}$ metallic states due to similar values between the bandwidth $W$ (corresponding to the kinetic hopping parameter $t$) and electronic correlation couplings (Hubbard repulsion $U$, Hund coupling $J_H$). In this case, the OSPP concept may be the key framework to understand the metallic behavior of this correlated system.

\textit{Method.}
In this work, the first-principles density functional theory (DFT) calculations were performed using the Vienna {\it ab initio} simulation package (VASP) code, and the projector augmented wave (PAW) method with the Perdew-Burke-Ernzerhof (PBE) exchange potential~\cite{Kresse:Prb,Kresse:Prb96,Blochl:Prb,Perdew:Prl}. Our plane-wave cutoff energy was $600$ eV. Furthermore, the $k$-point mesh was appropriately modified for different structures to render the in-plane $k$-point densities approximately the same in reciprocal space (e.g., $16\times8\times1$ for the monolayer-dimer phase). Those $k$-point meshes were tested to confirm that converged energies were produced. For the monolayers, a large vacuum ($\sim$20.0 \AA) was considered, to avoid interactions between layers. Both the in-plane lattice constants and atomic positions were fully relaxed until the Hellman-Feynman force on each atom was smaller than $0.01$ eV/{\AA}. Moreover, the rotationally invariant local spin density approach (LSDA) plus $U$ method was considered by using the Liechtenstein formulation with double-counting item~\cite{Liechtenstein:prb}. All the crystal structures were visualized with the VESTA code~\cite{Momma:vesta}. In addition to the standard DFT calculation discussed thus far, the maximally localized Wannier functions (MLWFs) method was employed to fit the Mo and Cr $3d$'s bands by using the WANNIER90 packages~\cite{Mostofi:cpc}. The two-dimensional Fermi surface was calculated by using the WannierTools software~\cite{WU2017}.

\textit{Orbital-selective Peierls transition in MoOCl$_2$.}
As shown in Fig.~\ref{Fig2}(a), the monolayer MoOCl$_2$~\cite{monocontext} contains MoO chains along the $a$-axis and MoCl$_2$ chains along the $b$-axis~\cite{DIMcontext}. For the dimerized phase of monolayer MoOCl$_2$, we obtained the optimized in-plane crystal constants $a = 3.775$~\AA ~and $b = 6.571$~\AA, in agreement with previous theoretical results ($a = 3.805$~\AA~and $b = 6.561$~\AA)~\cite{Zhao:prb20}. Based on our optimized crystal structure for the dimerized phase, the Mo-Mo dimer distortion is strong with $d_l$/$d_s$ $\sim 1.35$, resulting in a strong overlap of $d_{xy}$ orbitals in a dimer. According to group theory analysis using the AMPLIMODES software~\cite{Orobengoa:jac,Perez-Mato:aca}, this spontaneous distortion mode arising from the undistorted to the dimerized
phases is the Y$^{1+}$ mode.

Next, let us focus on the electronic structures for the non-magnetic (NM) state. It should be noted that the $d_{xy}$ orbitals are on the $bc$ plane, with the $x$- or $y$-axis along the Mo-Cl directions, while the $z$-axis is also the $a$-axis. As shown in Fig.~\ref{Fig2}(b), in the undistorted phase, the Mo $d_{xy}$ band displays a strongly quasi-one-dimensional electronic behavior along the $b$-axis, indicating Peierls instability along this direction, where the band structure is much more dispersive along the $b$-axis (X-S or Y-$\Gamma$ paths) than the $a$-axis ($\Gamma$-X or S-Y paths). However, in the dimerized phase, the Mo $d_{xy}$ band becomes flat with bonding-antibonding characteristics and opening a large gap $\sim 2.9$ eV, supporting the Peierls transition picture. In addition, the $d_{xz/yz}$ orbitals show itinerant properties with strong anisotropy along the $a$-axis, leading to metallic states in the $d_{xz/yz}$ sector. In this case, this system overall displays the OSPP metallic characteristics, confirming our intuitive analysis.

\begin{figure}
\centering
\includegraphics[width=0.48\textwidth]{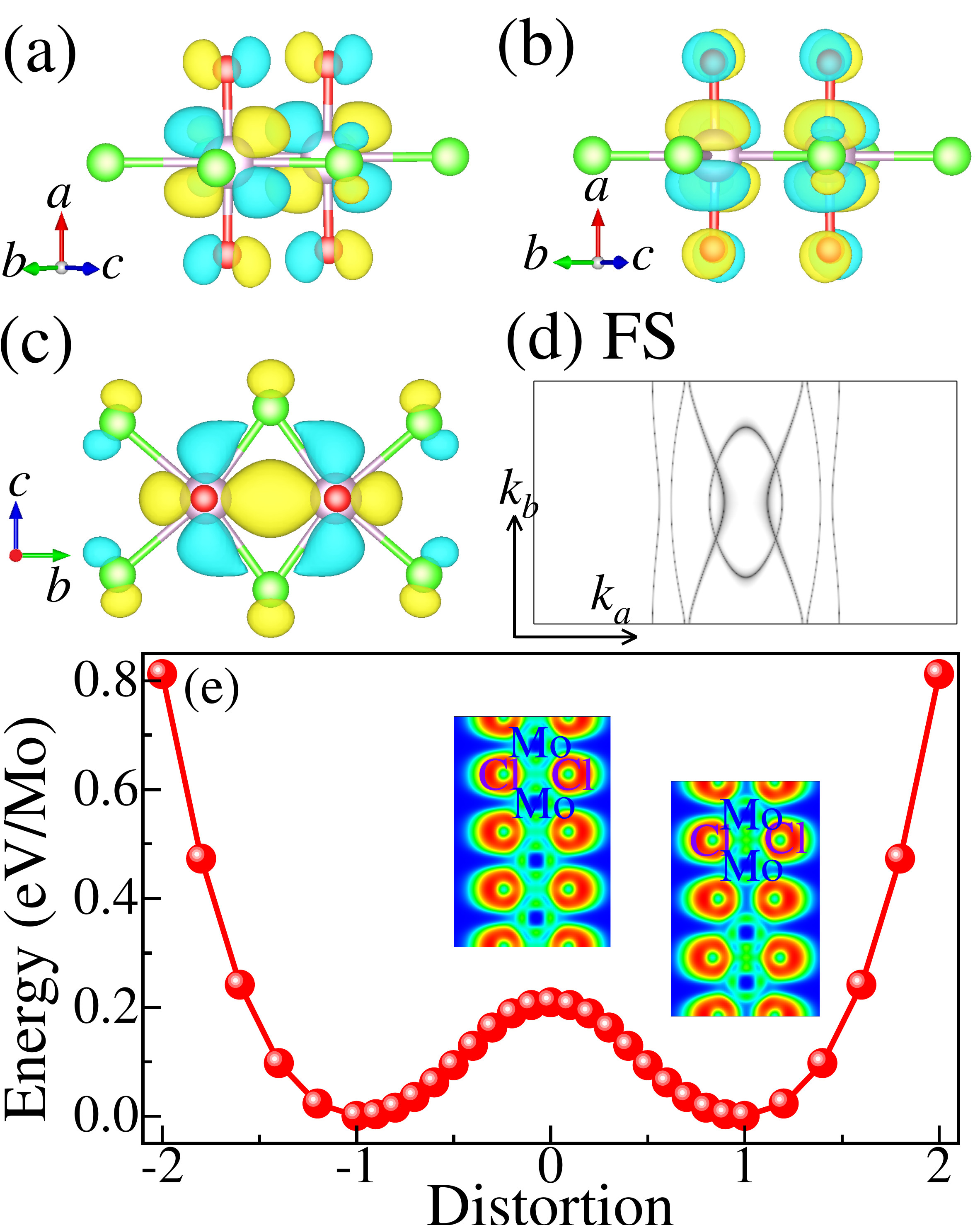}
\caption{(a-c) Wannier function of the Mo $t_{2g}$ orbitals in the dimerized phase. (a) $d_{xz}$ orbital. (b) $d_{yz}$ orbital. (c) $d_{xy}$ orbital. It illustrates the alternating $dd$$\sigma$-bonding and $dd$$\sigma^{*}$-antibonding states along the $b$-axis. (d) Fermi surface. (e) Calculated energy barriers for the switching path from the undistorted to dimerized phases, based on a simple linear interpolation of the Y$^{\rm 1+}$ mode by fixing the crystal constants of the dimerized phase. Inset:  Electron localization function for the monolayer MoOCl$_2$ corresponding to the undistorted and dimerized phases, in the $a-b$ plane.}
\label{Fig3}
\end{figure}

Based on the MLWFs method~\cite{Mostofi:cpc}, the disentangled Wannier functions of the Mo $t_{2g}$ orbitals are shown in Figs.~\ref{Fig3}(a-c). Clearly, the MoOCl$_2$ $d_{xy}$ orbitals form singlet dimers with a strongly $dd$$\sigma$-bonding state along the $b$-chain direction, whereas the $d_{xz}$ and $d_{yz}$ orbitals display different behavior. Based on the Wannier fitting (Fig. S1), we found that the hoppings in a Mo-Mo dimer are $t^{\vec{b}}_{xy-xy}$ $\sim 1.43$ eV and $t^{\vec{b}}_{xz/yz-xz/yz}$ $\sim 0.26$ eV. This large $d_{xy}$ orbital intra dimer hopping corresponds to a large bonding-antibonding $d_{xy}$ splitting (2$t^{\vec{b}}_{xy-xy}$ $\sim 2.9$ eV). In this case, $t^{\vec{b}}_{xy-xy}$ is larger than typical Hund couplings $J_H$ for $4d$ systems, often in the range $0.5 \sim 0.7$ eV~\cite{JH:prb}, indicating a strong molecular-orbital state in the $d_{xy}$ orbital dimer~\cite{Streltsovt:prb14}. Furthermore, the nearest neighbour (NN) inter-dimer $xy-xy$ hopping is about $0.09$ eV, which can be neglected compared with the intradimer $xy-xy$ hopping ($\sim$1.43 eV), leading to nearly ``decoupled'' singlet dimers.

Figure~\ref{Fig3}(d) displays the Fermi surfaces of the dimerized phase, made of the $d_{xz}$ and $d_{yz}$ orbitals. It shows a weak-band dispersion along the $b$-axis, suggesting highly anisotropic electron transport characteristics. This can explain the magnetoresistance behavior~\cite{Wang:prm20}, as discussed in Ref.~\cite{Zhao:prb20}. Furthermore, the NN hopping along the $a-$axis for the $d_{xz/yz}$ orbitals is $\sim 0.66$~eV, corresponding to a large bandwidth along the $a$-axis ($W \sim 4t^{\vec{a}} = 2.64 $ eV), in agreement with the strong itinerant character of the $d_{xz/yz}$ orbitals. In addition, the NN hopping for the $d_{xy}$ orbital is $\sim 0.02$ eV along the $a$-axis. Considering these $t_{2g}$ orbitals hopping parameters, the $d_{xy}$ states display molecular-orbital behavior, resulting in the Peierls transition along the $b$-axis, while the $d_{xz/yz}$ orbitals  show strong itinerant characteristics, leading to a strong anisotropy along the $a$-axis. Thus, the OSPP state is indeed the key concept to understand the metallic behavior of this system.

To better understand the Mo-Mo dimers along the $b$-axis, we simulated the switching path from the state without dimerization to the Mo-dimerized phase by a simple linear interpolation of the Y$^{\rm 1+}$ mode and fixing the crystal constants of the dimerized phase, as shown in Fig.~\ref{Fig3}(e). The transition energy barriers between those two phases is calculated to be $\sim 209$ meV/Mo, indicating this Mo-dimerization is quite stable.  In addition, we also calculated the electron
localization function (ELF)~\cite{Savin:Angewandte} for the undistorted and dimerized phases [inset in Fig.~\ref{Fig3}(e)]. After the Mo-dimerization distortion, the charges were more localized inside the Mo-Mo dimer than between dimers, resulting in a $dd$$\sigma$ bonding state.

\textit{Electronic correlations.}
Although there is no magnetic order in this material down to $1.8$~K, the results of magnetization measurements indicate the high-temperature magnetic susceptibility in the paramagnetic state contains $S \textless 1$ spins (compatible with the $S = 1$ for Mo$^{\rm 4+}$)~\cite{Wang:prm20}. The $d_{xy}$ orbital forms strong spin-singlet states ($(|{\uparrow \downarrow}\rangle -|{\downarrow \uparrow}\rangle)/\sqrt{2}$) in a dimer, while the decoupled orbitals can form both antiferromagnetic (AFM) and ferromagnetic (FM) arrangements. However, long-range order is difficult due to the nearly decoupled dimers. In this system, the quantum fluctuations are important to clarify the true magnetic ground state properties. Because DFT neglects fluctuations, we only calculated AFM and FM spin configurations within each Mo-Mo dimer~\cite{NMUcontext}, to analyze the nature of the orbital-selective Peierls metallic phase. Here, the strong intra-atomic interaction is introduced in a screened Hartree-Fock like manner, as the LSDA+$U$ method with Liechtenstein format within the double-counting item~\cite{Liechtenstein:prb}. In addition, we used $J_H = 0.6$ eV and $U = 3$ eV, considered to be suitable for typical $4d$ systems~\cite{Streltsovt:prb14,Korotin:sr,JH:prb}.

\begin{figure}
\centering
\includegraphics[width=0.48\textwidth]{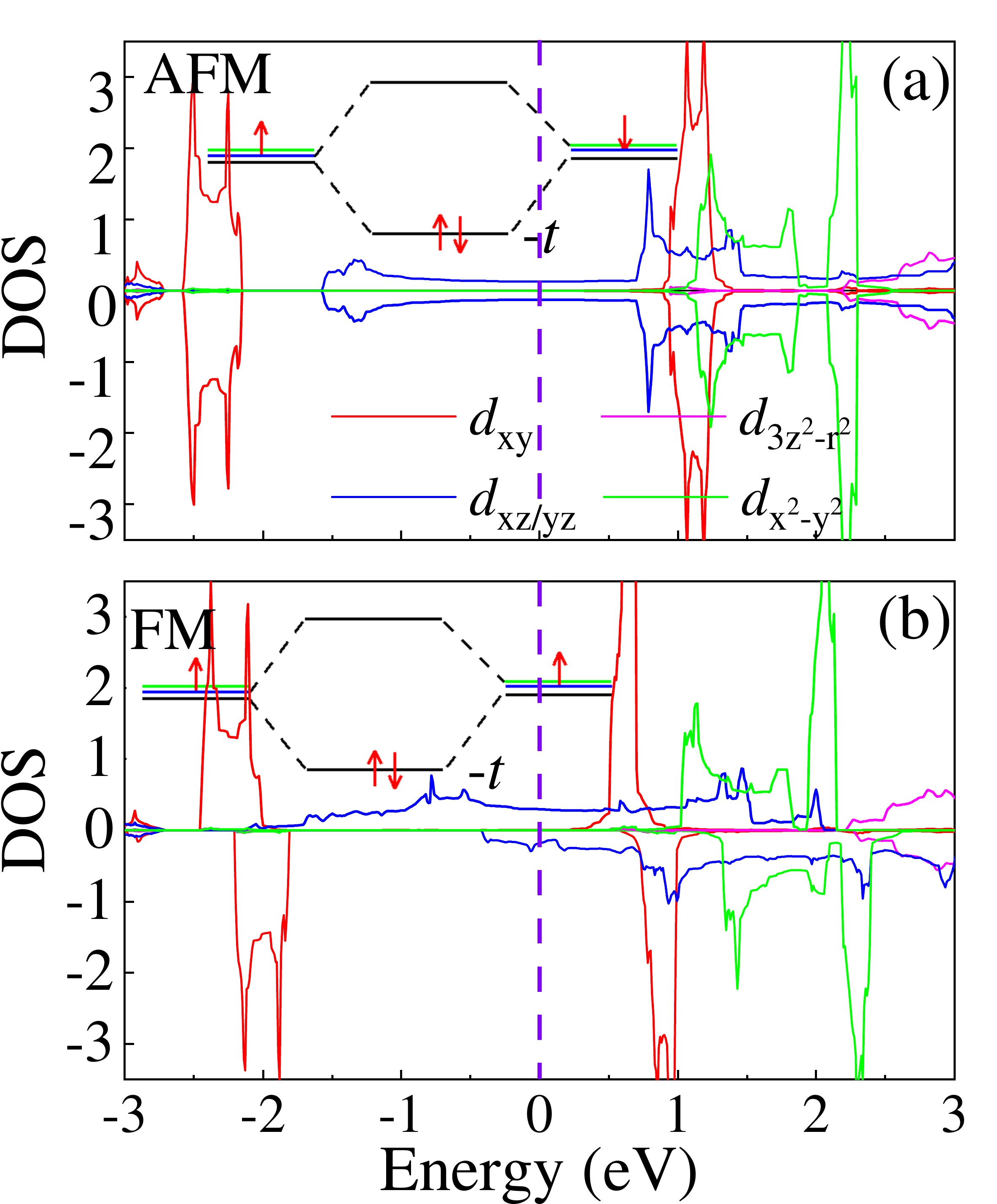}
\caption{(a) Projected density-of-states of the Mo-$d$ orbitals of monolayer MoOCl$_2$ with Mo-Mo dimerization for two spin configurations. (a) AFM arrangement of $d_{xz/yz}$ and spin-singlet state of $d_{xy}$. (b) FM arrangement of $d_{xz/yz}$ and spin-singlet state of $d_{xy}$. The five Mo $4d$ orbitals are distinguished by different colors. We used LSDA+$U$ with Liechtenstein format ($U = 3$ eV and $J_H = 0.6$ eV) on Mo sites.}
\label{Fig4}
\end{figure}

As shown in Fig.~\ref{Fig4}, the Mo $d_{xy}$ state displays strong bonding-antibonding characteristics in both AFM and FM dimers, indicating spin-singlet states ($(|{\uparrow \downarrow}\rangle -|{\downarrow \uparrow}\rangle)/\sqrt{2}$) for the $d_{xy}$ orbital. In addition, the other two orbitals ($d_{xz}$ and $d_{xz}$) are decoupled. Therefore, even considering reasonable $4d$ electronic correlation effects and magnetic coupling in a dimer, the $d_{xy}$ orbital still forms strong molecular orbitals, while the orbitals $d_{xz}$ and $d_{yz}$ provide the metallic conductive behavior of the ensemble.

Furthermore, the calculated local magnetic moment are $1.4$ $\mu_{\rm B}$/Mo and $0.81$ $\mu_{\rm B}$/Mo for AFM and FM arrangements, respectively. The reduced magnetic moment may be explained by the orbital-selective formation of covalent Mo–Mo bonds corresponding to orbital-selective behavior, as discussed in Ref.~\cite{Streltsov:pnas}. It should be noted that our main conclusion about a MoOCl$_2$ monolayer -- the formation of an orbital-selective singlet dimer and orbital-selective Peierls metallic phase -- can be expected to carry over to the bulk system. Detailed DFT results for the bulk MoOCl$_2$, namely for the full 3D system including the weakly coupled layers, can be found in the supplementary materials ~\cite{Supplemental}. The results are similar to the results of a monolayer MoOCl$_2$, confirming that this is essentially a two-dimensional material.

\textit{Discussion.}
As discussed in Ref.~\cite{Streltsovt:prb14}, molecular orbitals would be suppressed if the Hund coupling $J_H$ were larger or comparable to the hopping $t$ for one orbital in a dimer. In the vdW family with $3d^1$ configuration (VOI$_2$), the hopping for the $d_{xy}$ orbital in dimers is $t \sim 0.65$ eV ~\cite{Zhang:prb21}. This $t$ is smaller (or comparable) than a typical $J_H$ of the $3d$ system ($J_H \sim 0.7 - 1.0$ eV) ~\cite{JH:prb}. Hence, molecular orbitals may be suppressed in the $3d$ family of atoms. Although to our knowledge there are no related multi-orbital materials reported in this family with $d$ TM atoms, we used Cr instead of Mo to qualitatively discuss the possibility of orbital-selective singlet dimer formation in such a system. Here, starting from the optimized lattice constants of MoOCl$_2$, we fully relaxed the in-plane lattice constants and atomic positions for a monolayer CrOCl$_2$ ($d^2$) in the dimerized phase.

\begin{figure}
\centering
\includegraphics[width=0.48\textwidth]{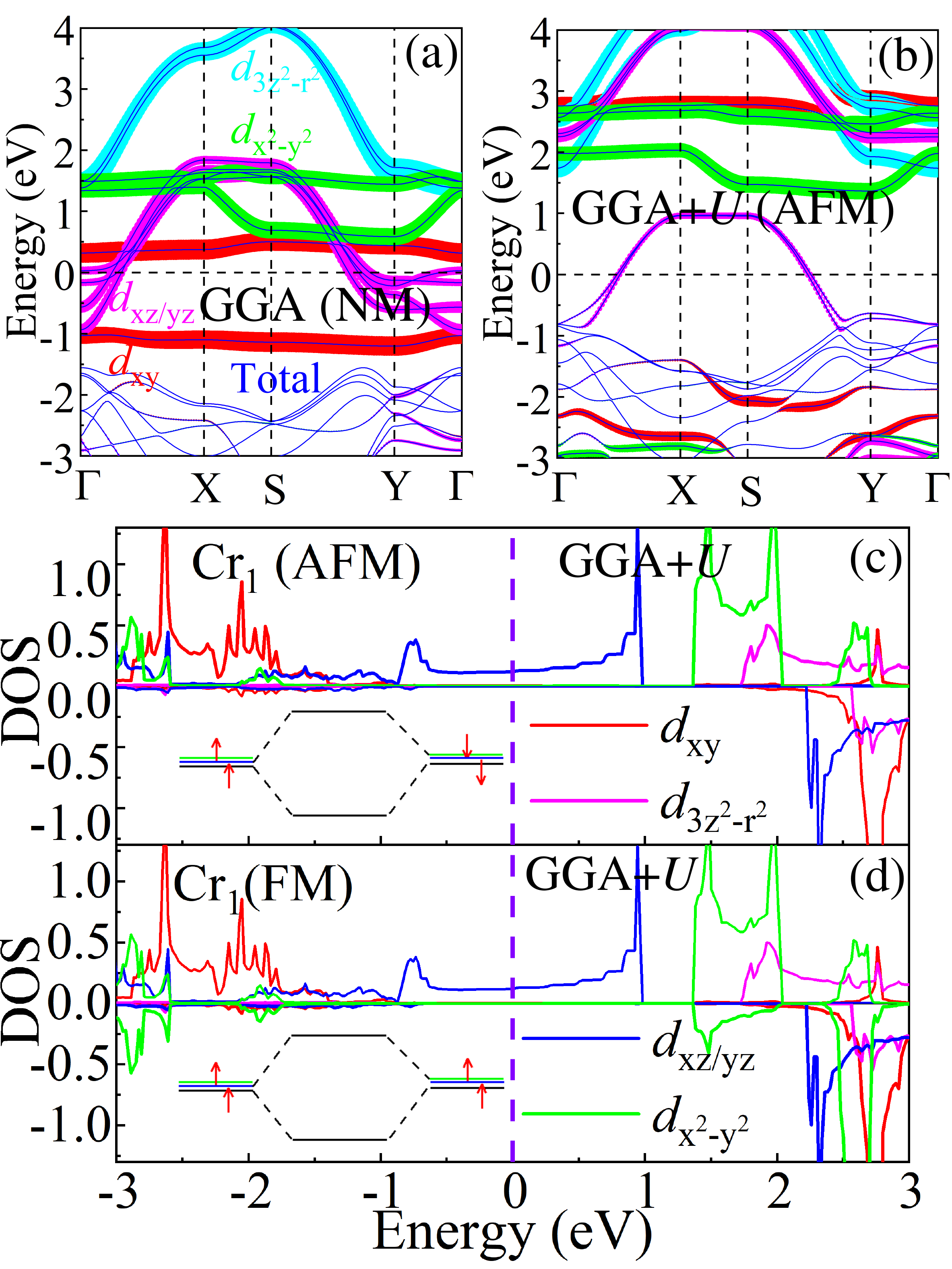}
\caption{(a-b) Projected band structures of the monolayer dimerized CrOCl$_2$ phase for (a) non-magnetic (GGA)  and (b) AFM (GGA+$U$) states, respectively. The weight of each Mo orbital is represented by the size of the circles. (c-d) Density-of-states of the Cr-$3d$ orbitals of monolayer CrOCl$_2$ with Cr-Cr dimer, projected for one Cr atom. (c) AFM state. (d) FM state. The five Mo $3d$ orbitals are distinguished by different colors. We used LSDA+$U$ with Liechtenstein format ($U = 4$ eV and $J_H = 0.8$ eV) on Cr site.}
\label{Fig5}
\end{figure}

In general, the $3d$ orbitals are less spatially extended than the $4d$ orbitals, leading to a reduced hopping $t$ in the $3d$ case. Furthermore, a typical $J_H$ of the $3d$ system is about $0.7-1.0$ eV, which is larger than in the $4d$ system ($J_H \sim$ 0.5-0.7 eV)~\cite{JH:prb}. Hence, in a $3d$ dimer system with multi orbitals, the singlet bonding molecular-orbital state may be destroyed due to an enhanced $J_H$ and a reduced $t$, as compared to the $4d$ system. As shown in Fig.~\ref{Fig5}(a), in the dimerized non-magnetic phase, the Cr $d_{xy}$ band forms the molecular orbitals with a gap $\sim 1.4$ eV, while the $d_{xz}$ and $d_{yz}$ orbitals display strong anisotropic behavior with metallic band characteristics. Based on Wannier functions fitting (Fig.~S1b), in the Cr-Cr dimer the intrahoppings are $t^{\vec{b}}_{xy-xy}$ $\sim 0.75$ eV and $t^{\vec{b}}_{xz/yz-xz/yz}$ $\sim 0.1$ eV. In this case, the hopping is comparable or smaller than the $J_H$ of a typical $3d$ system~\cite{JH:prb}. Then, the molecular-orbital formation and associated orbital-selective spin singlet should be suppressed by electronic correlations.

To qualitatively understand this issue, we calculated the AFM and FM spin arrangements in a dimer, based on the LSDA+$U$ with Liechtenstein format ($U = 4$ eV and $J_H = 0.8$ eV). As displayed in Fig.~\ref{Fig5}(b), the molecular orbitals of the Cr $3d_{xy}$ orbital are suppressed by electronic correlation. In addition, the $d_{xy}$ states display the Mott-insulating transition behavior, as shown in Figs.~\ref{Fig5}(c) and (d). Furthermore, the $d_{xz}$ and $d_{yz}$ states still display metallic behavior, resulting in a possible orbital-selective Mott (not Peierls) phase in this system. Because the TM-TM dimerization should be destroyed in this $3d$ system, long-range magnetic ordering or other interesting states could be stabilized. In addition, we calculated the phononic dispersion spectrum for the dimerized phase of the monolayer CrOCl$_2$. This clearly indicates the system is dynamically unstable (see Fig. S8). Thus, not only from the perspective in this publication
of comparing hoppings and Hund couplings of $3d$ and $4d$ systems, but from the phononic spectrum the
Cr-based dimerized system is unstable.
This issue deserves further theoretical and experimental studies. Hence, the OSPP stabilization seems difficult in the multiorbital $3d$ family in general.

\textit{Conclusions.-}
Here, based on DFT and DFT+$U$ calculations, we discussed a van der Waals layered family of materials. We clarified that the Mo-dimerized chain originates from the orbital-selective Peierls transition of the $d_{xy}$ orbital along the $b$-axis. In addition, the metallic behavior of MoOCl$_2$ is explained by the orbital-selective nature, where $d_{xy}$ forms insulating dimers while the other Mo-$d_{xz/yz}$ orbitals contribute to the metallic conductivity. Furthermore, due to the orbital-selective singlet behavior, leading to covalent bonds, the magnetic moments are reduced, as observed experimentally in this system. Finally, we qualitatively discussed a similar $d^2$ system but with $3d$ atoms. In this case, the orbital-selective Peierls state is suppressed by the enhanced electronic correlations and reduced intradimer hopping. Our results successfully reproduce the metallic behavior in some $4d$ orbital-selective dimerized systems and provide additional insight that should motivate further theoretical and experimental efforts.

\acknowledgments{The work of Y.Z., L.-F.L., A.M. and E.D. is supported by the U.S. Department of Energy (DOE), Office of Science, Basic Energy Sciences (BES), Materials Sciences and Engineering Division. All the calculations were carried out at the Advanced Computing Facility (ACF) of the University of Tennessee Knoxville (UTK).}


\begin{references}
\bibitem{Q3context} First, due to the octahedral crystal field, the two electrons occupied the three lower-energy $t_{2g}$ orbitals ($d_{xy}$, $d_{yz}$ and $d_{xz}$). Then, the energies of the $d_{yz}$ and $d_{xz}$ orbitals would shift down compared with the energy of the $d_{xy}$ orbital by considering the Jahn Teller distortion with $Q_3 \textless 0$.
\bibitem{monocontext} For two-dimensional vdW layered systems, the physical properties are mainly contributed by the Van der Waals layer. Here, to better fit the band structure and deduce the hoppings, we focus on discussing the monolayer.
\bibitem{DIMcontext} Both undistorted and dimerized monolayer MoOCl$_2$ are in an orthorhombic crystal structure with space group Pmmm (No.47), where the MoOCl$_2$ octahedra form a two-dimensional plane. However, in the undistorted phase, the MoOCl$_2$ chain has identical Mo-Cl bonds along the $b$-axis, while there are two different Mo-Cl bonds in the dimerized phase. Accordingly, the Cl atoms occupy two different Wyckoff sites ($2s$ and $2t$) in the dimerized state while the Wyckoff sites of Cl atoms are in $2s$ in the undistorted phase, corresponding to different in-plane symmetries. Due to the electronic occupation of the $d_{xz/yz}$ in this $d^2$ system, the ferroelectric distortion along $a$-axis is energetically unfavorable, as discussed in the electron doping for VOI$_2$ with $d^1$ configuration in Ref.~\cite{Zhang:prb21}.
\end{references}
\end{document}